\documentclass[12pt]{JHEP}
\usepackage{amsmath,amsfonts,amssymb}
 \usepackage{epsf}

\newcommand{\be}{\begin{equation}}
\newcommand{\ee}{\end{equation}}
\newcommand{\bea}{\begin{eqnarray}}
\newcommand{\eea}{\end{eqnarray}}

\newcommand{\bra}{{\langle}}
\newcommand{\ket}{{\rangle}}
\newcommand{\tr}{\hbox{ Tr}}
\newcommand{\anti}{{^\circ_\circ}}

\author{Samuel E. V\'azquez$^{1,\dagger}$\\
$^1$ Department of Physics, UCSB, Santa Barbara, CA 93106 \\
 $^\dagger$ \email{svazquez@physics.ucsb.edu} }

 \newcommand{\myfig}[3]{\begin{figure}[ht]
\begin{center}
\leavevmode \epsfxsize=#2cm \epsfbox{#1}
\end{center}
\caption{#3} \label{fig:#1}
\end{figure}}

\title{ BPS Condensates, Matrix Models and Emergent String Theory}
\abstract{ A  prescription is given for computing anomalous
dimensions of single trace operators in SYM at strong coupling and
large $N$ using a reduced model of matrix quantum mechanics. The
method involves treating some parts of the operators as ``BPS
condensates" which, in certain limit, have a dual description as
null geodesics on the $S^5$. In the gauge theory, the condensate is
similar to a representative of the chiral ring and it is described
by a background of commuting matrices. Excitations around these
condensates correspond to excitations around this background and
take the form of ``string bits" which are dual to the ``giant
magnons"  of Hofman and Maldacena. In fact, the matrix model
approach gives a {\it quantum} description of these string
configurations and explains why the infinite momentum limit
suppresses the quantum effects. This method allows, not only to
derive part of the classical sigma model Hamiltonian of the dual
string (in the infinite momentum limit), but also its quantum
canonical structure. Therefore, it provides an alternative method of
testing the AdS/CFT correspondence without the need of
integrability. }

 \preprint{}

\begin{document}

\section{Introduction}

Understanding the strong coupling limit of non-abelian gauge
theories is still an outstanding  open problem in theoretical
physics. Most of our understanding comes from Conformal Field
Theories (CFTs). According to the AdS/CFT conjecture, at large $N$
and large but fixed 't Hooft coupling, we should find an effective
geometrical description of these theories in terms of perturbative
string theory on an asymptotically AdS background \cite{adscft}.
Proving (or disproving) this conjecture is still an important open
problem. However, much evidence in its favor have been found in
recent years. The best studied example of the correspondence is the
duality between ${\cal N} = 4$ SYM theory in four dimensions and
string theory on $AdS_5\times S^5$. This is the model that we will
study in this paper.

The most significant obstruction to proving the correspondence is
our inability to make strong coupling calculations in SYM theory.
Pure supersymmetric states have protected dimensions and so one can
map them directly to SUGRA excitations on the dual string
theory\footnote{See \cite{review} for a more comprehensive review of
this and other aspects of the AdS/CFT correspondence.}. However, for
non-supersymmetric states one needs to be more clever and find a way
to extrapolate a weak coupling calculation to strong coupling. As in
any perturbative expansion, the secret is to be careful in choosing
the ``background" state one is expanding around.

It was realized a few years ago that one can define the perturbation
theory around the supersymmetric single trace 1/2 BPS states of the
form \be \tr(Z^J)\;,\ee with $Z$ one of the three complex adjoint
scalars of SYM. For large $J$ these states are dual to point-like
classical strings rotating with angular momentum $J$ along a null
geodesic on the equator of the $S^5$. The first ``small" non-BPS
excitations to be understood where the so-called BMN states of the
form \cite{bmn} \be \label{BMNstate} \sum_l e^{i p l} \tr(Z^{J-l} X
Z^l X)\;,\ee where $p \sim 1/J$.

 In fact, the effective geometry seen by the small strings is a
plane wave and thus it allows the exact calculation of the spectrum
in the string theory side. It turns out that for these fast rotating
strings the expansion parameter in their energy is of the form $\sim
\lambda /J^2$. It is then possible to extrapolate a perturbative
calculation in SYM by {\it first} taking $J$ large and then $\lambda
\rightarrow \infty$ but with $\lambda/J^2 = \text{fixed} \ll 1$.

On the string theory side, however, the limit is the opposite:
$\lambda \gg 1$ {\it first} to get perturbative string theory, and
then focus on trajectories with $J \gg 1$. Surprisingly, both
calculations agree up to order $\lambda^3$ where a $1/J$ discrepancy
is has been found \cite{nearbmn}. It is believed that the
discrepancy is just an artifact of the opposite order of limits and
the fact that keeping $\lambda /J^2$ fixed entangles both
$1/\sqrt{\lambda}$ and $1/J$ quantum corrections.

One can proceed along these lines and consider states with many
transverse ``impurities", e.g. \be \tr(Z X X Z X \cdots)\;.\ee One
can compute the anomalous dimension matrix in this basis and it has
been found that one can also define large charge limits with $J_X,
J_Z \rightarrow \infty$ that allow to extrapolate the weak coupling
calculation to strong coupling. On the string theory dual one find
that these states correspond to long semiclassical rotating
strings\footnote{There is an extensive literature on this subject so
here we will refer the reader to the recent review
\cite{stringreview}.}. As before, one considers a limit where
$\lambda/ J^2$ is held fixed. One can even match the sigma model
actions for these strings in this limit \cite{kru1, kru2, kru3}, but
one again finds a discrepancy at three loops \cite{minahan1}.

One can also define an expansion around BPS D-brane states called
Giant Gravitons. The limits described above are also useful for this
case and one can show agreement with the string theory at one loop
\cite{dds1, dds3}.

It is apparent at this point that all of the comparisons described
above require certain amount of confidence in extrapolating the weak
coupling results in SYM to strong coupling. One would like to be
able to define a strong coupling expansion directly. This is of
course the ``holy grail" of high energy physics. In any case, it is
believed that this expansion is possible at least in SYM based on
{\it integrability}.

In the last years there has been great hope that integrability will
allow to prove AdS/CFT at least in the free closed string sector.
This hope was motivated by the discovery of integrability at one
loop in the gauge theory dilatation operator \cite{minahan} and was
followed by the discovery of {\it classical} integrability in the
string theory world-sheet \cite{polchinski}.

The quantum integrability in the gauge theory has been argued to
persist at higher loops and an all-loop guess for the Bethe ansatz
has been presented in \cite{beisert1, beisert2}. This has also been
accompanied by a similar guess for the presumed quantum Bethe ansatz
for the dual string theory \cite{Arutyonov}. The all loop Bethe
ansatz for the gauge theory was actually used in the higher loop
test of AdS/CFT described above\footnote{For a recent review about
integrability and semiclassical strings see \cite{plefkareview}.}.

One can see that the proposed integrable structures are very similar
on both sides of the correspondence but they do not quite agree. In
fact, all the disagreement can be encoded in a single overall phase
in the Bethe ansatz' $S$-matrix \cite{Arutyonov}. Solving this
discrepancy is still an open problem. In any case, using
integrability to prove/test AdS/CFT has some major drawbacks. First,
integrable structures are very delicate and can fail at higher loop
as it happens in the plane wave matrix model \cite{Klose}. Failure
of integrability at two loops has also been claimed to happen for
the open strings on Giant Gravitons \cite{openint,davidsam}. This is
specially concerning since all of the test of AdS/CFT so far rely on
an (educated) guess at higher loops. In fact, the quantum
integrability of the string theory is still a conjecture. Secondly,
even if integrability works for ${\cal N} = 4$ SYM it will not work
for other less symmetric CFTs.

It is then very desirable to develop techniques that allow a
systematic large $\lambda$ expansion directly from the gauge theory
without the need of integrability. We believe that the key for this
program lies again in expanding around highly supersymmetric states.
Moreover, given the general combinatorial nature of the dilatation
operator in terms of spin chains (at least for the scalar sector),
it seems very likely that one can always find a reduced quantum
mechanical model to describe the effective dynamics of certain
sectors of the theory. In fact this idea was originally put forward
by Berenstein in \cite{toymodel}. It was argued there that the
effective dynamics of 1/2 BPS states can be described by a reduced
quantum mechanical model of a single matrix in an harmonic
oscillator potential which, after diagonalization, can be written as
$N$ fermions in an harmonic trap. One can describe semiclassical
states of the theory in terms of droplets in the single particle
phase space. Reduced quantum mechanical sectors are also known to
arise in thermal ${\cal N} = 4$ SYM on $\mathbb{R} \times S^3$
\cite{Yamada,Harmark}.

For the 1/2 BPS states, the reduced model was amazingly confirmed by
a SUGRA calculation in \cite{LLM}, where it was found that all the
1/2 BPS solutions in IIB SUGRA can also be classified in terms of
droplets in a plane.

Generalizations of the reduced matrix model for 1/4 and 1/8 BPS
states were proposed in \cite{berens} in terms of multiple matrix
models of commuting matrices. The gravity side of the story for
these states is still incomplete, but some recent progress has been
made in \cite{Donos1}. Nevertheless, some important consistency
checks for the proposal in \cite{berens} have been put forward
recently. First, the ground state for the 1/8 BPS model is described
by a singular distribution of eigenvalues on an $S^5 \subset
\mathbb{R}^6$ in the large $N$ limit \cite{berens}. One can
calculate quadratic excitations around this background of commuting
matrices and one finds that they can be pictured as ``string bits"
which joint two eigenvalues on the sphere. The spectrum of these
excitations has a dispersion relation \cite{dds2}, \be
\label{dispersion} E(p) = \sqrt{1 + \frac{\lambda}{\pi^2}
\sin^2(p/2)} = \sqrt{1 + \frac{\lambda}{\pi^2} \sin^2(\Delta \varphi
/2)}\;,\ee where $p$ is the momentum of the BMN state
(\ref{BMNstate}) and $\Delta \varphi$ is the azimuthal angle between
the two eigenvalues on the sphere. This is in fact the exact
proposed dispersion relation in the Bethe ansatz of both the gauge
theory and the string dual \cite{beisert1,beisert2,Arutyonov}.

What is more surprising is that the picture of the ``string bits"
and the relation $p = \Delta \varphi$ was recently confirmed by
Hofman and Maldacena  using a purely classical string theory
analysis \cite{diegomalda}. These authors introduced a limit where
one takes $J \rightarrow \infty$ with $p$ fixed. It was proposed
that the ``magnon" excitations of the form (\ref{BMNstate}) are dual
to classical string solutions on $S^2$ that form a straight line
joining two points on the boundary of the circular droplet (the
equator of $S^2$ in the usual coordinates). These ``giant magnons"
have a dispersion relation like (\ref{dispersion}) (at large
$\lambda$) after identifying $p = \Delta \varphi$. This confirms the
prediction of the matrix model calculation! Moreover, these matrix
model techniques can also be applied to more general CFTs as in
\cite{diego, davidcota}.

Therefore, even though we do not have a {\it proof} that the
effective dynamics of scalar operators is given by a reduced matrix
model, we have non-trivial checks that this is indeed the case.

In this article we attempt to clarify the meaning of the matrix
model calculations. In particular we give a precise proposal that
relates the matrix model computations to the more familiar operator
mixing problem. This is done in terms of what we call ``BPS
condensates" which are summarized in section 2. A detailed
discussion is given in sections 3 and 4 where we also review how one
obtains the giant magnons of \cite{diegomalda} directly from the
matrix model calculation and how one can generalize these to the
$SU(3)$ sector of the theory. For the $SU(3)$ sector, however, the
matching with the dual string theory is more restricted since it
turns out that one needs to understand the backreaction to the 1/4
and 1/8 BPS condensates.

In any case, we obtain a quantum description of the giant magnons
and we get a better understanding of why the Hofman-Maldacena limit
is really a classical limit. The possible interpretation of magnon
bound states in our formalism is briefly discussed in section 3.  We
also show how the matrix model calculation gives not only the
correct Hamiltonian for the string states (in the infinite momentum
limit) but also the canonical structure expected from the string
theory dual. This is very encouraging since it would be very
desirable to match directly the sigma model of the string theory and
its canonical structure rather than having to solve for its
spectrum. This would allow generalizations to other less symmetric
field theories. In section 5 we discuss the backreaction to the BPS
condensates and we explain why our method works in the strong
coupling limit. Finally we discuss the prospects to relate this
procedure to the more familiar Bethe Ansatz technique on the
conclusion.

\section{The General Idea of BPS Condensates} Here we want to
summarize the general idea of BPS condensates using the familiar
single trace scalar operators of SYM theory. In the next sections we
will develop the details of this method and its interpretation in
terms of the dual string theory.

When doing perturbative calculations in either the gauge theory or
the string dual one always has to choose a classical background
configuration to expand upon. In the gauge theory side of the
correspondence one can rephrase this as defining the expansion
around certain set of operators that are dual to classical string
configurations as we discussed in the introduction.

Here we want to rephrase this expansion in terms of the effective
matrix model mentioned above. Lets start by considering a generic
$SU(2)$ single trace operator. One can write this kind of state
using the bosonized language introduced in \cite{dds1, dds3}: \be
\label{su2excited} \tr(Y Z^{n_1} Y Z^{n_2} Y\cdots )\;.\ee In the
next section we argue  that these states are described by a quantum
mechanical matrix model of two complex matrices. For large $n_i$ one
can see the $Y$s as impurities in an otherwise 1/2 BPS operator
$\tr(Z^n)$. In the excited state (\ref{su2excited}), the $Z^{n_i}$s
look localy like BPS states and we call them {\it 1/2 BPS
condensates}. In the matrix model it then makes sense to expand
around a background of normal matrices: $[Z, \bar Z] = 0$ and $Y =
0$. This ``classical" BPS background can be the circular droplet for
example and we identify it with the 1/2 BPS condensate in the
operator language. The fluctuations $\delta Y$ are called  ``string
bits" and they are dual to commutators between the $Y$ and the BPS
condensates in the operator language: \be \tr(Z^n) \rightarrow
\tr([Y,Z^{n_1}] [Y,Z^{n_2}]\cdots)\;,\;\;\; \sum_i n_i = n\;.\ee The
states (\ref{su2excited}) serve as a basis for these excitations.
The fluctuation $\delta Z$ are the backreaction of the condensate
and in this case we will see that they can be integrated out to give
an effective action for the transverse excitations on the classical
BPS background.

In the dual string theory we will see that for $n_i \rightarrow
\infty$ the 1/2 BPS condensates become classical and localize the
ends of the string bits on the boundary of the circular droplet.
Namely, a null geodesic on $\mathbb{R} \times S^1$.  This is
precisely the interpretation advocated recently in
\cite{diegomalda}. This will be discussed in detail in the next
section.

 If the number of
$Y$ fields is comparable to the number of $Z$s,  it makes sense to
expand around a state of the form: \be \tr(\{ Z^n Y^m\})\;.\ee Here
the curly brackets denote symmetrization between the letters $Y$ and
$Z$. This state is just a rotation of the 1/2 BPS state. For
multi-trace operators, symmetrized states similar to this one are
1/4 BPS \cite{Ryzhov1,Ryzhov2}. Small excitations around this state
will be described by turning on commutators that break the
symmetrization, \be \tr(\{ Z^n Y^m\}) \rightarrow \tr([Y,\{Z^{n_1}
Y^{m_1}\}] [Z,\{Z^{n_2} Y^{m_2}\}] \cdots)\;,\;\; \text{etc.}\ee

 One
can find a basis for these excitations in analogy to
(\ref{su2excited}), \be \tr(Y \{Z^{n_1} Y^{m_1}\} Z \{Z^{n_2}
Y^{m_2}\} \cdots)\;.\ee  Since the words $ \{Z^{n_i} Y^{m_i}\}$ look
like  1/4 BPS states we call them {\it 1/4 BPS condensates}. As in
the case of the 1/2 BPS condensates this description is more useful
when $n_i, m_i \rightarrow \infty$. As we will see, the dual
interpretation is that, in this limit, the 1/4 BPS condensates
localize the ends of the string bits on null geodesic on $\mathbb{R}
\times S^3$.

 In the matrix model language we should then expand
around a ``classical" configuration of commuting normal matrices
$[Y,Z]=0 = [Z, \bar Z] = [Y , \bar Y] = 0$. This configuration can
be one of the 1/4 ``droplets" of \cite{berens}. For the ground state
we get a $S^3$ distribution of eigenvalues. The fluctuations around
this background will correspond to the commutators: \be \delta Z
\simeq [Z,\{ Z^{n_i} Y^{m_i}\}]\;,\;\;\; \delta Y \simeq [Y,\{
Z^{n_i} Y^{m_i}\}]\;.\ee We discuss these fluctuations in section 4.

The generalization to $SU(3)$ states should be obvious by now. For
example, in the case where we have many $Y$ and $Z$ fields and a few
$X$s, we can consider the following basis for the transverse
excitations \be \label{su3basis1} \tr(X \{ Z^{n_1} Y^{m_1}\} X \{
Z^{n_2} Y^{m_2}\}\cdots )\;.\ee In the matrix model we expand around
$[Y,Z]= [Y, \bar Y] = [Z, \bar Z] = 0$, $X = 0$. In this case a
fluctuation $\delta Y$ or $\delta Z$ amounts to a breaking one the
condensates and thus leaving this restricted basis, e.g. \be \{
Z^{n_i} Y^{m_i} \} \rightarrow [Z,\{ Z^{n_i - 1} Y^{m_i} \}] \;.\ee
On the other hand, a $\delta X$ fluctuation is just a commutator
between an $X$ and one of the condensates.

We can now try to consider states with many $Z$s (for example) and
similar quantities of $Y$ and $X$. In this case it make sense to
expand in a basis like \be \tr(X Z^{n_1} Y Z^{n_2} \cdots)\;.\ee In
the matrix model we diagonalize $Z$ and expand around $[Z, \bar Z] =
0$, $X= Y = 0$ just like in the $SU(2)$ case.

Finally we can have 1/8 BPS condensates by considering symmetrized
combinations $\{ X^{n_i} Y^{m_i} Z^{p_i}\}$. In this case we expand
around configurations with three normal commuting matrices. The
fluctuations are dual to the commutators between any of the fields
and the condensates just like for the $SU(2)$ states.

\section{1/2 BPS Condensates: The $SU(2)$ Sector}

In this section we briefly review the 1/2 BPS states and their
effective dynamics in terms of a normal matrix model. We then
consider generic $SU(2)$ states an set up an effective description
in terms of a two matrix model similar to the one in \cite{dds2}. We
then construct the Hilbert space in terms of the 1/2 BPS
condensates. We obtain the canonical commutator relations for these
states and explain how to obtain their classical limit. In doing so,
we recover the picture of the ``string bits" of \cite{berens,dds3}
and explain its precise relation with the ``giant magnons" of
\cite{diegomalda}. Moreover, we match the canonical structure
expected from the dual string theory and also its sigma model in the
limit of large $\lambda$ and infinite angular momentum. Finally, we
comment on the a possible interpretation of the bound states of
\cite{diegomalda, Dorey1, Dorey2, Minahan1} in terms of the matrix
model.

\subsection{Review of 1/2 BPS  States Dynamics}
The 1/2 BPS states of SYM theory are described by multitrace
operator build out of a single complex scalar: ${\cal O}_{n_1 n_2
\cdots} = \tr(Z^{n_1}) \tr(Z^{n_2})\cdots$ (see \cite{Corley} and
references therein). They are eigenstates of the dilatation operator
with dimension\footnote{In this paper we will make heavy use of the
operator/state correspondence of SYM. We will go back and forth
between operators and states and we hope that the context will make
clear which one we are using.} \be \label{eigen} \hat \Gamma |{\cal
O}_{n_1 n_2 \cdots} \ket = (n_1 + n_2+ \ldots)|{\cal O}_{n_1 n_2
\cdots} \ket = \hat J_z |{\cal O}_{n_1 n_2 \cdots} \ket\;. \ee

It is well known that at one loop, the contribution from the D-term
in the scalar potential $V_D\sim\tr|[Z,\bar Z]|^2$ to the dilatation
operator is canceled by fermion and gauge loops \cite{Constable}.
One can translate this to the dual Hilbert space as, \be
\label{normal} \bra {\cal O}_{n_1 n_2 \cdots} | \tr|[Z,\bar Z]|^2
|{\cal O}_{n_1' n_2' \cdots}\ket = 0\;. \ee Since these states are
protected by supersymmetry, we expect this to be true independently
of the gauge coupling.

From Eqs. (\ref{eigen}) and (\ref{normal}) it is natural to guess
that the effective dynamics of these states will be described by a
normal gauged matrix model with an harmonic oscillator potential
\cite{toymodel, Corley, Takayama, Ghodsi}, \be \label{bpsaction} S =
\int dt \tr (|D_t Z|^2 + |Z|^2)\;, \;\;\;\; [Z, \bar Z] = 0\;.\ee
This model can be visualized as a reduction of SYM on $\mathbb{R}
\times S^3$ down to the zero mode of a single scalar \cite{berens,
toymodel}. The eigenstates of the matrix model can be expressed as
antisymmetric wave function of the complex eigenvalues $z_i$ and can
be classified in terms of Young Tableux \cite{toymodel}. Their
quantum numbers match (\ref{eigen}).

Moreover, in the large $N$ limit generic coherent states are
described by ``droplet"-like distributions of eigenvalues on the
complex plane. For example, the ground state $\psi_0 \sim e^{-\tr
(Z\bar Z)}$ will have a probability density, \be \bra \psi_0 |
\psi_0 \ket  = \int_{[Z,\bar Z] = 0} [dZ d\bar Z]| \psi_0|^2 \propto
\int \prod_{i = 1}^N d^2 z_i\,\, e^{-2 \sum_i |z_i|^2 + \sum_{i< j}
\log|z_i - z_j|^2 } \;,\ee
 which will be
dominated by the saddle point of the exponential in the large $N$
limit. Here we have used the measure change for a normal matrix
model which follows from the metric $ds^2 = \tr(dZ d\bar Z)$
\cite{matrixmodels}.

In the limit $N \rightarrow \infty$ one replaces the sums by density
distributions and one extremizes the functional \be E[\rho] = -2
\int d^2 z \rho(z) |z|^2 + \frac{1}{2} \int \int d^2 z_1 d^2z_2
\rho(z_1) \rho(z_2) \log|z_1 - z_2|^2 \;\;,\ee with the constraint
$N = \int d^2 z \rho(z)$. Using the fact that the logarithm is the
Green's function in two dimensions one obtains a circular droplet
distribution of eigenvalues of constant density given by
\cite{matrixmodels} \be \sigma = - \frac{\Delta W(z,\bar z)}{4 \pi}
= - \frac{\partial \bar \partial W(z,\bar z)}{\pi} =
\frac{2}{\pi}\;,\ee where $W(z,\bar z) = -2 |z|^2$ is the potential
for the eigenvalues. Then, from the normalization of the density one
obtains the radius of the droplet: $r_0 = \sqrt{N/2}$. The droplet
approximation is very useful for calculating correlation functions
in position space as we will see below.

We would like to comment that there is a possible second matrix
model that gives the same description of the 1/2 BPS states. This is
a {\it complex} matrix model with the action (\ref{bpsaction}). It
turns out that one can transform $Z$ as $Z \rightarrow U^\dagger
(Z_\text{diag} + R) U$, where $U \in SU(N)$, $Z_\text{diag}$ is
diagonal and $R$ is strictly upper triangular \cite{matrixmodels}.
The measure change is \cite{matrixmodels}, \be [dZ d\bar Z] \propto
 \prod_{i =1}^N d^2 z_i \prod_{j < k } |z_j - z_k|^2 \prod_{m < n} d^2 R_{m n}\;.\ee
One can easily see that the ground state wave function will be the
same as in the normal matrix model and  that for {\it holomorphic}
correlation functions, the matrix $R$ does not contribute
\cite{matrixmodels}. Therefore,  the dynamics of this model (in the
holomorphic sector) is the same as with a normal matrix. However, as
we will see,  the distinction can be important for multiple matrix
models.

\subsection{$SU(2)$ States Dynamics}
A generic $SU(2)$ operator has the form: \be \label{14bpsstate}
\tr(Z Y Z Z \cdots )\tr(Z Y Y Z \cdots )\cdots\;. \ee For these
holomorphic states one still has that the D-term contributions add
to zero at one loop. Therefore, we expect that we can also ignore
the D-terms from the effective matrix model. The lack of
supersymmetry can make this model much more complicated that the 1/2
BPS case. However one can argue along the lines of \cite{berens}
that for the 1/4 BPS states the corresponding matrix model is a
simple rotation of (\ref{bpsaction}): \be \label{su3bpsaction}
S_{\text 1/4 BPS} = \sum_{\alpha = Y,Z} \tr\left( |D_t Z_\alpha|^2 +
|Z_\alpha|^2 \right)\;, \ee for normal commuting matrices
$[Z_\alpha, Z_\beta] = [Z_\alpha, \bar Z_\alpha] = 0$.

At one loop, the anomalous dimension is again generated by the
commutators from the F-terms: $\tr|[Z_\alpha, Z_\beta]|^2$. We can
again argue for an all-loop generalization \cite{dds2} involving two
{\it complex} matrices and ignoring the D-terms: \be \label{action}
S = \int dt \tr\left[ \sum_\alpha (|D_t Z_\alpha|^2 + |Z_\alpha|^2 )
+ \frac{g_{YM}^2}{(2 \pi)^2} \sum_{\alpha, \beta} |[Z_\alpha,
Z_\beta]|^2 + \text{higher commutators} \right] \;. \ee The first
three terms of the action come from the direct reduction of SYM on
the $S^3$ to the zero mode of the matrices, and the higher
commutators will come from integrating out higher modes and
fields\footnote{Here the $SO(6)$ invariant potential of SYM can be
written as $\sum_{a, b = 1}^6 \tr|[X_a, X_b]|^2 = 2 \sum_{\alpha,
\beta = 1}^3 \tr|[Z_\alpha, Z_\beta]|^2 + 2 \sum_{\alpha, \beta =
1}^3 \tr|[Z_\alpha, \bar Z_\beta]|^2$, where $Z = \frac{1}{\sqrt 2}
(X_1 + i X_2)$ etc. The last term in the potential is the D-term
that we ignore in this matrix model.}. A similar matrix model arises
from the one-loop dilatation operator in the $SU(2)$ sector
\cite{Agarwal,Bellucci}.

As in any quantum system one chooses a particular classical
configuration for which to define the perturbation theory. To define
a consistent perturbative expansion one needs to find stable
classical configurations. In our case, we know that in the large $N$
limit (and with $Y = 0$) the normal matrix model $[Z, \bar Z] = 0$
can be described by droplets on the single particle phase space of
the eigenvalues of $Z$. Moreover, these configurations are
stabilized by SUSY. It then makes sense to expand around these
``classical" solutions. Since the classicality is only statistical,
one needs a prescription to define this expansion. We do this in the
following way. First write as usual $Z \rightarrow Z  + \delta Z$,
and we diagonalize the background $Z$. Then,
 we treat the eigenvalues $z_i$ as random numbers with probability
 distribution $\sim \exp(-2 \sum_i |z_i|^2 + \sum_{i < j} \log |z_i
 - z_j|^2)$. The resulting operators and states of the Hilbert space
 will depend on $z_i$. Therefore, we define the inner product to be
 the statistical average of the usual one:
 \be
 \bra \phi| \hat{\cal O} |\tilde\phi\ket \equiv \frac{\int \prod_i d^2 z_i
  |\psi_0(\{z_i\})|^2 \bra \phi(\{z_i\})| \hat{\cal O}(\{z_i\}) |\tilde\phi(\{z_i\})\ket}{  \int \prod_i d^2 z_i
  |\psi_0(\{z_i\})|^2 } \;.\ee
  In the large $N$ limit we can use the saddle point approximation
  as before.

  We now need to take into account the backreaction to 1/2 BPS
  background, $\delta Z$. If we ignore the possible higher
  commutators in the action (\ref{action}), we see that the
  fluctuation $\delta Z$ enters quadratically in the action and can
  be integrated out leaving an effective action for the matrix $Y$
  in the BPS background $Z$. This will result in higher interactions
  for $Y$. In this article we will ignore these interactions for
  simplicity. These higher interactions
  are suppressed in the strong coupling and infinite momentum limit
  as we discuss below.

 At this point, our procedure  is pretty much
  equivalent to the one in \cite{Donos,Rodrigues}, but the approach presented here
  is simpler and allows generalizations beyond the $SU(2)$ sector as
  we will see in the next sections. Matrix models in terms of
  collective fields  describing the BMN limit of SYM have also been
  studied in \cite{deMelloKoch1,deMelloKoch2}.

  Finally, one also has the possibility of expanding
  around the ground state of the {\it complex } matrix model
  mentioned in the previous section. It turns out that the
  distinction between the two backgrounds goes to zero in the
  infinite momentum limit discussed in the next section.

  \subsection{1/2 BPS Condensates, String Bits and Giant Magnons}

  We can now calculate the effective hamiltonian for the $Y$ fields on the background of $Z$. After diagonalizing the
  normal matrix $Z$, we have,
\be \label{HY} H = \sum_{i , j} \omega_{i j} \left[
(A^\dagger_Y)_i^j (A_Y)^i_j + (A^\dagger_{\bar Y})_i^j (A_{\bar
Y})^i_j\right]  + \text{interactions} \;,\ee where the creation
operators are given by \bea (A_Y^\dagger)_i^j &=& \sqrt{\frac{w_{i
j}}{2}} [Y_i^j - \frac{i}{\sqrt{w_{i j}}} (\pi_Y)_i^j]\;,
\\ (  A_{\bar Y}^\dagger)_i^j &=& \sqrt{\frac{w_{i j}}{2}} [
\bar Y_i^j -  \frac{i}{\sqrt{w_{i j}}} (  \pi_{\bar Y})_i^j]\;, \eea
with \be \omega_{i j} = \sqrt{1 + \frac{\lambda}{(2 \pi)^2} |z_i -
z_j|^2}\;.\ee Here we have assumed that the fermions cancel the
zero-point energy from normal ordering the operators in (\ref{HY}).
Finally, we have normalized the diagonal matrix so that $z_i  = r_i
e^{i \phi_i}$ with $r_i \leq 1$.

We can now construct the Hilbert space of states. With our
normalization a generic single trace state takes the form \be
\label{bosonized} | n_1, n_2,\ldots, n_L\ket =
\frac{1}{N^{L/2}}\tr(A_Y^\dagger \psi_{n_1}(Z) A_Y^\dagger
\psi_{n_2}(Z) \cdots A_Y^\dagger \psi_{n_L}(Z) ) |0\ket_Y \;, \ee
where, \be \psi_n(Z) = \sqrt{1+ n} Z^n \;,\ee and $|0\ket_Y$ is the
usual vacuum for $Y$ defined by $A_Y |0\ket_Y = A_{\bar Y} |0\ket_Y
= 0$. The wavefunctions $\psi_n(Z)$ are dual to the  1/2 BPS
condensates. In fact as we will see they will localize on an $S^1$
at large $n$ just like the 1/2 BPS states.  This is the infinite
momentum limit. The $A_Y^\dagger$ excitations are called ``string
bits" \cite{berens,dds2} and as we will see they have a dual
description as the ``giant magnons" of \cite{diegomalda}.

It is easy to verify the orthonormality of these states using our
inner product prescription: \be \label{inner} \bra n_1, n_2, \ldots,
n_L | n_1', n_2',\ldots, n_L'\ket \approx \prod_{l = 1}^L \int
[dD_l] \psi_{n_l}(z_l)^* \psi_{n_l'}(z_l)  = \prod_{l = 1}^L
\delta_{n_l, n_l'}\;,\ee where $\int [dD] = \int_0^1 dr r
\int_0^{2\pi} d\phi/\pi$ is the integration across the droplet, and
we are assuming the generic case where not all of the $n_i$ are
equal so we ignore the cyclicity of the trace.

Now lets consider calculating the expectation value of some
observable, say the Hamiltonian (\ref{HY}). After doing the usual
planar contractions of the $A_Y$s one can always reduce the problem
to a product of integrals over the droplet. A useful property of the
1/2 BPS condensates is \be Z \psi_n(Z) = \sqrt{\frac{n+1}{n+2}}
\psi_{n+1}(Z)\;.\ee Under the inner product (\ref{inner}) one then
sees that $Z$ and $\bar Z$ can be treated as the operators  $\hat
Z^\dagger$ and $\hat Z$ with the property: \bea \label{12bpsop1}
\hat Z^\dagger |n \ket &=& \sqrt{\frac{n + 1}{n+2}} |n+1\ket \equiv
\hat r e^{i \hat \phi} |n\ket\\
\label{12bpsop2}
 \hat Z |n \ket &=&
\sqrt{\frac{n}{n+1}} |n-1\ket \equiv  e^{-i
\hat \phi}  \hat r|n\ket\;, \\
 \eea
 where we define
 \be \hat r = \sqrt{\frac{\hat n}{\hat n + 1}}\;,\;\;\; e^{i \hat \phi} |n
 \ket = |n+ 1\ket\;.\ee

 Note that $\bra \hat r\ket \leq 1$ and thus as the notation suggests, this
 will become the operator that measures the radial distance of the
 droplet. We also observe that $\hat n$ is the momentum conjugate to
 $\hat \phi$,  $ [ \hat \phi , \hat n]  = i$, or in terms of $\hat r$, $\hat p_\phi  = \hat r^2/(1 - \hat r^2)$.
 From the canonical commutator
 relation we can derive,
 \be \label{canonical1} [ \hat \phi, \hat r ] = i \frac{(1 - \hat r^2)^2}{2 \hat r}\;.
 \ee
This indicates that our system is constrained, which is not
surprising since we are restricting our Hilbert space by choosing
these special $SU(2)$ states. As we will see, this is exactly the
expected canonical structure for the $SU(2)$ states in the string
theory dual after an appropriate gauge choice.

It is easy to show that for any function $f(Z,\bar Z)$ with a power
law expansion, \be \int [dD] \psi_n(z)^* f(z,\bar z) \psi_{n'}(z)
\cong \bra n| \anti f(\hat Z^\dagger, \hat Z) \anti |n'\ket\;,\ee
where $\anti \anti$ denotes anti-normal ordering with respect to the
operators $\hat Z^\dagger$, $\hat Z$. Therefore, we can write the
quadratic Hamiltonian in our basis as \be \label{H2su2} H^{(2)} =
\sum_{l=1}^L \anti \sqrt{1 + \frac{\lambda}{(2\pi)^2} |\hat
Z^\dagger_l - \hat Z^\dagger_{l+1}|^2} \; \anti\;.\ee

Let us now discuss the possibility of expanding around the
background corresponding to the ground state of the {\it complex}
matrix model discussed in the previous section. Ignoring again the
backreaction, we regard the matrix $Z$ as a random variable with
distribution $\int [dZ d\bar Z] \exp(-2 \tr(Z\bar Z))$. Now consider
the Hamiltonian at one-loop in $\lambda$. The anomalous dimension
will be generated by a term $ \propto \tr[Z ,A_Y^\dagger][\bar Z,
A_Y] $.

 Lets apply this to the state $|\psi \ket \sim \tr(\cdots
Z^{n_1} A_Y^\dagger Z^{n_2} \cdots)$. It is easy to see that the
anomalous dimension Hamiltonian will act in such a way to introduce
a $Z$ and $\bar Z$ fields at the left and right of $A_Y^\dagger$ so
that we conserve the $U(1)$ charge. As an example, consider the
interaction that takes $Z^{n_1} \rightarrow Z^{n_1}  \bar Z  Z$.
Then, the matrix elements for this kind of interactions will be
reduced to the following correlation function, \be \frac{ \bra
\tr(\bar Z^{n_1'} Z^{n_1} \bar Z  Z) \ket}{ \sqrt{ \bra \tr(\bar
Z^{n_1'} Z^{n_1'}) \ket \tr(\bar Z^{n_1} Z^{n_1}) \ket}} \approx
\frac{N}{2} \delta_{n_1, n_1'}\;, \ee where we have taken the planar
limit.

Now consider doing the same correlation function but with the normal
matrix model (NMM), \be \frac{ \bra \tr(\bar Z^{n_1'} Z^{n_1} \bar Z
Z) \ket_\text{NMM}}{\sqrt{ \bra \tr(\bar Z^{n_1'} Z^{n_1'})
\ket_\text{NMM} \bra \tr(\bar Z^{n_1} Z^{n_1}) \ket_\text{NMM}}}
\approx \frac{\int [dD] \bar z^{n_1' + 1} z^{n_1+1}}{\int [dD]
|z|^{2 n_1} }  =  \frac{N}{2} \left(\frac{n_1 + 1}{n_1 + 2}\right)
\delta_{n_1, n_1'}\;.\ee We see that both results agree only in the
infinite spin limit $n_1 \rightarrow \infty$.

Therefore,  we can regard the calculations done in this paper with a
normal matrix background as asymptotic in the large spin limit. In
any case, we can calculate corrections to this limit with the
backreaction term $\delta Z$.  We leave these calculations for the
interested reader.

\subsection{Classical Limit and the dual String Theory}

In Quantum Mechanics one usually recovers the classical limit by
taking $\hbar \rightarrow 0$. This makes the canonical commutators
vanish, $[x, p] = i \hbar \rightarrow 0$ so that $x$ and $p$ become
simple classical observables.  For constrained systems the classical
limit can be trickier. In our case we can see from
(\ref{canonical1}) that the classical limit is reached by taking
states for which $\bra \hat r \ket \rightarrow 1$, or $\bra \hat n
\ket \rightarrow \infty$\footnote{Note that in our conventions
$\hbar = 1$.}. This is precisely the localization on the edge of the
droplet which is correlated to the limit where the $Y$ impurities
are ``far away". This limit also takes away the ordering ambiguities
from the Hamiltonian which becomes \be \label{angleH2} H^{(2)}
\approx \sum_{l=1}^L \frac{\sqrt \lambda}{\pi} \sin
\left(\frac{\phi_l - \phi_{l+1}}{2}\right) \;,\ee where we have
taken the large $\lambda$ limit to compare with the string theory.

 This is precisely the Hamiltonian of the ``giant
magnons" of \cite{diegomalda}. In fact, the whole picture is exactly
the same: we can picture the $Y_i^j$ as an excitation (string bit)
joining two eigenvalues $z_i$ and $z_j$ which become localized at
the edge of the droplet when $n \rightarrow \infty$. In fact, taking
$n\rightarrow \infty$ with $\Delta \phi$ fixed is precisely the
Hoffman-Maldacena limit.  Here, however, we have a full quantum
description of the system. One can see that finite $n$ effects will
delocalize the ends of the string bits and make them ``fuzzy" on the
$S^2$ (fig. 1). \myfig{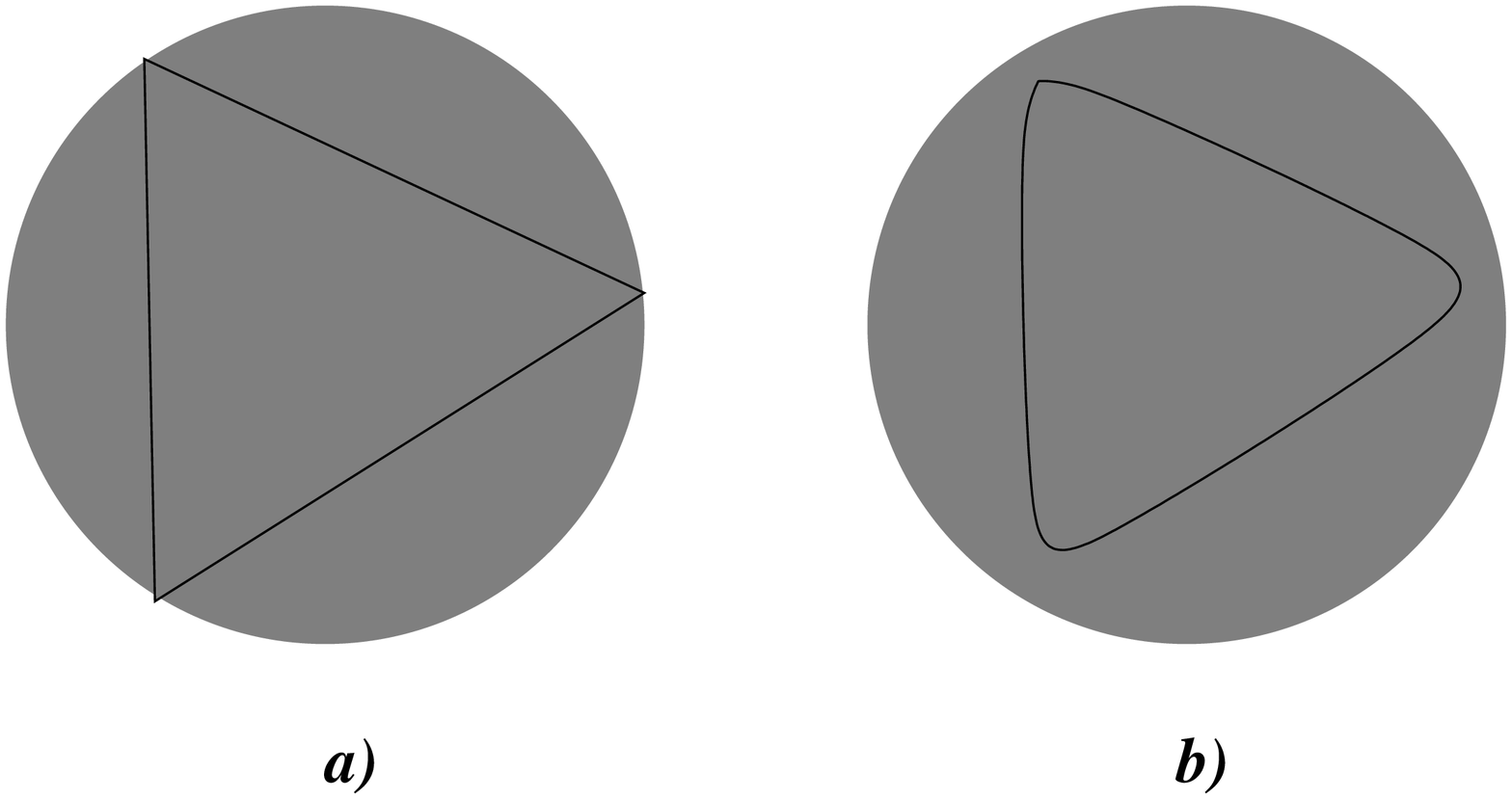}{10}{In figure a) we show the usual
giant magnons at infinite momenta $n_i \rightarrow \infty$. The
lines are identified with the ``string bits" $Y_i^j$ and the edges
with the 1/2 BPS condensates $Z^{n_i}$. Here the disk is mapped to
an $S^2 \subset S^5$ so that the edge is the equator of the $S^2$.
Finite $n_i$ effects will delocalize the ends of the magnons away
from the equator as shown in figure b).} Finite $n$ effects were
considered recently in \cite{Arutyunov:2006gs} for a single giant
magnon. Their results confirm that the ends of the giant magnon are
delocalized from the equator at finite $n$.

The relation between the momentum of the magnons and the angles on
the droplet can be seen more clearly by Fourier transforming as in
\cite{dds2}.  For example, consider the following asymptotic state,
\be \label{asympstate} |\psi(p)\ket = \lim_{\Delta \rightarrow
\infty} \lim_{n \rightarrow \infty} \sum_{x = -\Delta}^\Delta e^{i p
x} \tr(\cdots Y^\dagger Z^{n + x} Y^\dagger Z^{n - x} Y^\dagger
\cdots) |0\ket_\text{Y}\;,\ee where we first take the limit of
infinite separation $n \rightarrow \infty$. One can show that these
states are approximately orthogonal in this limit.

For condensates of infinite angular momentum the operators
(\ref{12bpsop1}) and (\ref{12bpsop2}) become simple shift
generators, \be \hat Z |n\ket \approx |n-1\ket\;,\;\;\; \hat
Z^\dagger |n\ket \approx |n+1\ket\;.\ee Focusing on a single string
bit, the effective Hamiltonian between these asymptotic states is
simply, \be H^{(2)}_{l, l+1} \approx \sqrt{1 +
\frac{\lambda}{(2\pi)^2} { \cal  H}_{l, l+1}}\;,\ee where, \be {\cal
H}_{l, l+1} |n_l, n_{l+1}\ket = 2|n_l,n_{l+1}\ket - |n_l - 1, n_l +
1\ket - |n_l + 1, n_{l+1} - 1\ket \;.\ee  This is the same
asymptotic Hamiltonian found in the two matrix model of
\cite{Rodrigues}. The eigenstates of this Hamiltonian are simply
plane waves of the form (\ref{asympstate}) with energy, \be
E_\text{asymp.} = \sqrt{1 + \frac{\lambda}{\pi^2}
\sin^2\left(\frac{p}{2}\right)}\;.\ee Comparing with (\ref{angleH2})
we see that $p  \cong \Delta \phi$.

We can also match the canonical structure of these string bits to
the one found in the string theory side. Since these are $SU(2)$
states they must be a limit of the well known rotating strings on
$\mathbb{R} \times S^3$ \cite{stringreview}. A well known limit of
these strings is the one corresponding to ``long strings": $L
\rightarrow \infty$ with $\lambda/L^2 = \text{fixed} \ll 1$ as
discussed in the introduction. The canonical structure should not
depend on the particular limit we are taking since $L$ is a
conserved quantum number. One can calculate the Polyakov action for
these string using the following coordinates on the $S^3$
\cite{dds3}: \be Z = r e^{i (t + \phi)}\;,\;\; Y = \sqrt{1 - r^2}
e^{i \varphi}\;.\ee

One then chooses the gauge, $\tau = t$, $p_{\varphi} =
\text{const.}$, which is appropriate to compare with the bosonized
labeling of the states (\ref{bosonized}). One obtains the
action\footnote{In \cite{dds3} we droped the factor of -1 since we
were comparing anomalous dimensions only.}, \be \label{stringaction}
S \approx L \int dt \int_0^1 d\sigma \left[\frac{r^2 \dot{\phi}}{1 -
r^2} -1 - \frac{ \lambda}{8 \pi^2 L^2} (r'^2 + r^2 \phi'^2) + {\cal
O}\left(\frac{\lambda^2}{L^4} \right)\right]\;, \ee where one
eliminates the time derivatives of all the higher order terms
\cite{kru2,kru3} and, \be L = \sqrt{\lambda} \int_0^1 d\sigma
p_\varphi\;,\;\;\; \lambda = g_{YM}^2 N  = R^4/\alpha'^2\;.\ee

The same action can be found from the spin chain formalism in the
gauge theory at one loop using the coherent states for the Cuntz
algebra \cite{dds3}. From this action we see that the canonical
momenta are \be p_r = 0\;,\;\;\; p_\phi  = L \frac{r^2}{1 -
r^2}\;.\ee This are precisely the constraints found above from the
matrix model calculation!

The additional factor of $L$ comes from taking the large $L$ limit,
since the total momentum for the $Z$ fields in the gauge theory
becomes: \be \hat J_Z = \sum_{l = 1}^L \hat n_l \approx L \int_0^1
d\sigma \frac{\hat r^2}{1 - \hat r^2}\;.\ee

One can also reproduce the commutator algebra by using the classical
Dirac brackets for constrained systems \cite{dirac}. The Dirac
bracket is defined by \be \{A, B\}_{\text{D}} = \{ A,
B\}_{\text{PB}} - \sum_{i,j} \{A, f_i \}_{\text{PB}} (G^{-1})_{i j }
\{f_j, B\}_{\text{PB}}\;,\ee where $\{,\}_{\text{PB}}$ is the usual
Poisson bracket. Furthermore, the second class constraints are given
by the equations $f_i = 0$ and $G^{-1}$ is the inverse of \be G_{i
j} = \{ f_i, f_j\}_{\text{PB}}\;.\ee The constraints are given by:
\be \label{constraints} f_1 = p_r\;,\;\;\; f_2 = p_\phi - L
\frac{r^2}{1 - r^2}\;.\ee Then it is straightforward to verify \be
\{\phi, r\}_{\text{D}} = \frac{(1 - r^2)^2}{2 L r}\;.\ee The
comparison with the quantum theory is done as usual: $[ , ] = i \{ ,
\}_{\text{class.}}$. One then obtains precisely the continuum
version of the commutators (\ref{canonical1}).

\subsection{Multiple Giant Magnons and Bound States}
Suppose that we put many string bits together as in the state
(\ref{su2excited}). One of two things can happen: either we have the
trivial addition of classical giant magnons, or we form a bound
state.  The first outcome happens if we take the formal $n_i
\rightarrow \infty$ limit for each condensate. The total energy in
the limit of many string bits ($L \gg 1$) is, \be H^{(2)}  \approx
\sum_{l=1}^L \frac{\sqrt \lambda}{\pi} \sin \left(\frac{\phi_l -
\phi_{l+1}}{2}\right) \approx \frac{\sqrt{\lambda}}{2\pi}\int_0^1
d\sigma
\partial_\sigma \phi = \frac{\sqrt{\lambda}}{2\pi} \Delta \phi\;.\ee
Here we assume that $L$ does not scale in any way with $\lambda$. If
we scale $L$ as in the fast string limit, $\lambda/L^2 \ll 1$ we get
instead, \be H^{(2)} \approx  L \int_0^1d\sigma\left( 1 +
\frac{\lambda}{ 8 \pi^2 L^2} \phi'^2 + \cdots\right)\;.\ee This
agrees with the classical string action (\ref{stringaction}) at one
loop.

 Understanding the emergence of bound states and of strings at $r < 1$,  requires taking into
account $1/n_i$ corrections. However, there are two effects that can
be important for understanding these corrections. First, as we
discussed above we need to take into account the backreaction term
in the matrix model. Furthermore, one needs to understand the
possible higher order interactions that can come from integrating
higher spherical harmonics in SYM. We do not have a good
understanding of these issues at this moment. However, suppose that
these unknown interactions tend to normal order the Hamiltonian
(\ref{H2su2}). Then one can deduce the classical limit by
constructing coherent states for the operators $\hat Z_l$.

In fact, one can show that the states, \be |z\ket = (1 -
|z|^2)\sum_{n = 0}^\infty \sqrt{n + 1} z^n |n\ket\;,\ee are indeed
overcomplete coherent states of the operator $\hat Z$. The
completeness relation is, \be \int_0^1 \frac{dr r}{(1 -
r^2)^2}\int_0^{2\pi} \frac{d\phi}{\pi} |z\ket \bra z| = 1\;.\ee The
classical Hamiltonian in the coherent state basis will be, \be \bra
: H^{(2)} : \ket = \sum_{l=1}^L  \sqrt{1 + \frac{\lambda}{(2\pi)^2}
| Z_l -  Z_{l+1}|^2}  \approx \frac{\sqrt\lambda}{2\pi}\int_0^1
\sqrt{r'^2 + r^2 \phi'^2} \;,\ee where in the last step we have
taken the strong coupling limit and then the continuum limit
corresponding to a large number of string bits. This is exactly the
Nambu-Goto action for a string on $\mathbb{R}\times S^2$ in the
static gauge \cite{diegomalda}.

Of course, the presence of bound states follows directly from the
duality between this action and the Sine-Gordon theory
\cite{diegomalda}. These bound states and its generalizations have
been studied recently in \cite{Dorey1, Dorey2,
Minahan1,new1,new2,new3}.

\section{1/4 and 1/8 BPS Condensates: The $SU(3)$ Sector}

The concept of 1/4 BPS condensates was discussed briefly in section
2. Here we will study these condensates in detail as we did for the
1/2 BPS case. The general states we want to consider are of the form
\be \label{generic14BPS}\tr(X \{ Z^{n_1} Y^{m_1}\} Y \{ Z^{n_2}
Y^{m_2}\} Z \cdots )\;.\ee In particular we are interested in the
limit where $n_i,  m_i \rightarrow \infty$ with the number of
``impurities" outside of the condensates held fixed.

By the same arguments of the previous section, we expect an
effective description of generic $SU(3)$ states in terms of a matrix
model similar to (\ref{action}) but now with three complex matrices
($\alpha = X, Y, Z$). There are two complications in the $SU(3)$
case, however. The fluctuations $\delta X$,  $\delta Y$ and  $\delta
Z$ cannot be integrated out so easily. This is actually a good thing
as these fluctuations have an interpretation in the operator
language as we discussed in section 2.

Moreover, it turns out that we need to include the D-terms to the
action. This conclusion follows from comparing with the string
theory results (see below). We do not have a purely field
theoretical explanation for this but it seems to be a consequence of
the fact that we have less supersymmetry for these states and hence
the one loop result is not protected at strong coupling.  The
effective action for these states is then of the form, \bea
\label{su3action} S &=& \int dt \tr\left[ \sum_\alpha (|D_t
Z_\alpha|^2 + |Z_\alpha|^2 ) + \frac{g_{YM}^2}{(2 \pi)^2}
\sum_{\alpha, \beta} |[Z_\alpha, Z_\beta]|^2 \right.  \nonumber \\
&& \left.  + \frac{g_{YM}^2}{(2 \pi)^2} \sum_{\alpha, \beta}
|[Z_\alpha, \bar Z_\beta]|^2 + \text{higher commutators} \right] \;.
\eea

As we discussed in section 2, for large $n_i, m_i$ the states
(\ref{generic14BPS}) look like many ``condensates" of 1/4 BPS
operators involving $Z$ and $Y$ fields.  Therefore we can try to
define our expansion around ``classical" configurations with $[Z, Y]
= [Z ,\bar Z] = [Y, \bar Y] =0$.  The classical configurations will
be the eigenstates of the 1/4 BPS action (\ref{su3bpsaction})  in
the large $N$ limit. In particular, we know that the ground state is
given by $\psi_0 \sim e^{-\tr(|Z|^2 + |Y|^2)}$ \cite{berens}. Thus,
the energy functional that determines the geometry of the eigenvalue
distribution is \be E[\rho] = -2 \int d^4 x \rho(x) (|z|^2 + |y|^2)
+ \frac{1}{2} \int \int d^4 x_1 d^4x_2 \rho(x_1) \rho(x_2) \log(|y_1
- y_2|^2 + |z_1 - z_2|^2) \;,\ee where we have used the measure
change for two normal commuting matrices: \be [dZ d\bar Z dY d\bar
Y] \propto \prod_i d^2 z_i d^2y_i \prod_{k < l} (|y_k - y_l|^2 +
|z_k - z_l|^2)\;.\ee

Doing a saddle point calculation as in \cite{dds2} one finds that
the eigenvalues form a singular distribution on an $S^3 \subset
\mathbb{R}^4$ with radius $r_0 = \sqrt{N}/2$. The fluctuations
around this background are the non-BPS parts of these operators.

The motivation for this interpretation is the same as with the 1/2
BPS condensates. If we consider calculating the one loop anomalous
dimension for these states we see that the planar contribution from
the F-term and D-term commutators $\tr|[Z, Y]|^2$, $\tr|[Z, \bar
Y]|^2$ will be zero when contracted between the condensates. Only
contractions between fields outside the condensates and the
condensates themselves will give rise to anomalous dimensions. We
interpret these interactions as fluctuations around the commuting
background. The interactions $\tr|[Z, X]|^2$, $\tr|[Z, \bar X]|^2$
and $\tr|[Y, X]|^2$, $\tr|[Y, \bar X]|^2$ that involve a condensate
and a transverse $X$ field are interpreted as the fluctuation
$\delta X$ in the matrix model. Finally the interactions $\tr|[Z,
Y]|^2$, $\tr|[Z, \bar Y]|^2$  with $Z$ or $Y$ fields outside the
condensate are just the backreaction of the condensate $\delta Y$
and $\delta Z$.

\subsection{Hilbert Space and Canonical Structure}

To simplify the discussion let us start with the states \be
\label{special14BPS} \tr(X \{ Z^{n_1} Y^{m_1}\} X \{ Z^{n_2}
Y^{m_2}\} X \cdots )\;.\ee That is, let us ignore the backreaction
to the 1/4 BPS condensates for the moment ($\delta Y = \delta Z =
0$). We then diagonalize the classical background of commuting
normal matrices. The effective Hamiltonian for the $X$ fluctuations
will have the same form as for the $SU(2)$ case (\ref{HY}) but now
with the following dispersion relation \be \omega_{i j} = \sqrt{1 +
\frac{\lambda}{(2 \pi)^2} \left(|y_i - y_j|^2 + |z_i -
z_j|^2\right)}\;.\ee

Note that if we had ignored the D-terms, the dispersion relation
would have an additional factor of 1/2: $\sqrt{1 + \frac{\lambda}{2
(2\pi)^2} (|\Delta z|^2 + |\Delta y|^2)}$.

In analogy with the 1/2 BPS condensates, the basis for this sector
is \be \label{su3bosonized} | n_1, m_1 ; n_2 , m_2 ; \ldots; n_L,
m_L\ket = \frac{1}{N^{L/2}}\tr[A_X^\dagger \psi_{n_1, m_1}(Y,Z)
\cdots A_X^\dagger \psi_{n_L,m_L}(Y,Z) ] |0\ket_X \;, \ee with \be
\psi_{n, m}(Y,Z) = \sqrt{\frac{(n + m + 1)!}{n! m!}}
  Z^n Y^m\;.\ee
Using the saddle point approximation, the inner product can be
reduced to integrals over an $S^3$ in the large $N$ limit: \bea
\label{inner} \bra n_1, m_1 ; \ldots; n_L, m_L | n_1', m_1'; \ldots;
n_L' ,m_L'\ket &\approx& \prod_{l = 1}^L \int
\frac{(d\Omega_3)_l}{\text{Vol}(S^3)} \psi_{n_l, m_l}(y_l, z_l)^*
\psi_{n_l',m_l'}(y_l,z_l)\nonumber \\
 &=& \prod_{l = 1}^L \delta_{n_l,
n_l'} \delta_{m_l,m_l'}\;,\eea where $z, y$ become coordinates on
the three-sphere, $|y|^2 + |z|^2 = 1$.

Just like in the $SU(2)$ sector, we can treat $Z$ and $Y$ as
creation operators under our inner product. They obey the following
properties:

 \bea \label{su3canonical} \hat Z^\dagger |\psi_{n,m}\ket  &=& \sqrt{ \frac{n
+1}{2 + n + m}} |\psi_{n+1,m}\ket \equiv \frac{\hat r_1}{\sqrt{1 +
\hat \eta}}
e^{i \hat \phi_1} |\psi_{n,m}\ket \;, \\
 \hat Y^\dagger |\psi_{n,m}\ket  &=& \sqrt{ \frac{m +1}{2 + n +
m}} |\psi_{n,m+1}\ket \equiv \frac{\hat r_2}{\sqrt{1 + \hat \eta}}
e^{i \hat \phi_2} |\psi_{n,m}\ket \;, \\
e^{i \hat \phi_1} |\psi_{n,m}\ket &=& |\psi_{n+1,m}\ket\;, \\
e^{i \hat \phi_2} |\psi_{n,m}\ket &=& |\psi_{n,m+1}\ket\;, \\
\hat r_1 &=& \sqrt{\frac{\hat{n}}{\hat n + \hat m}} \;, \\
\hat r_2 &=& \sqrt{\frac{\hat{m}}{\hat n + \hat m}}\;,\\
\hat r_1^2  + \hat r_2^2 &=& 1 \;, \\
 \hat\eta &=& \frac{1}{\hat n + \hat m}\;.\eea
We can now calculate the following canonical structure: \bea
\label{su3commutator} [\hat \phi_1, \hat r_1] &=& i \frac{\hat
r_2^2}{2 \hat r_1} \hat \eta
\;,\\
 {[}\hat \phi_1, \hat r_2 {]} &=& -i \frac{\hat r_2}{2} \hat \eta \;,\\
 \label{su3commutator3} {[}\hat \eta, \hat \phi_\alpha {]}  &=& i \hat \eta^2 \;,\eea and the
conjugate momentum to $\hat\phi_\alpha$ is $\hat P_{\alpha} = \hat
r_\alpha^2 /\hat\eta = \hat n_\alpha$.

One can also show that the operators are antinormal ordered under
the inner product (see Appendix A) and so the effective quadratic
Hamiltonian is just \be \label{su3Ham} H^{(2)} = \sum_{l=1}^L \anti
\sqrt{1 + \frac{\lambda}{(2\pi)^2} \left(|\hat Y^\dagger_l - \hat
Y^\dagger_{l+1}|^2 + |\hat Z^\dagger_l - \hat Z^\dagger_{l+1}|^2
\right)} \; \anti\;.\ee

\subsection{Localization and the Classical Limit}
The localization of the 1/4 BPS condensates works just like for the
1/2 BPS case. Looking at the canonical commutators
(\ref{su3canonical}) we see that for states with $\bra \hat n + \hat
m \ket \rightarrow \infty$  with $\bra \hat n \ket /\bra \hat m\ket$
 fixed will localize on an $S^3$. The operators $\hat r_\alpha$,
 $\hat \phi_\alpha$ become commuting (classical) numbers and we can drop the anti-normal ordering
 symbols on the  Hamiltonian (\ref{su3Ham})  and replace the operators by classical
 coordinates on the $S^3$, $|Y_l|^2 + |Z_l|^2 = 1$.

 In the case of the 1/2 BPS condensates, the localization occurs
 from $S^2 \rightarrow S^1$. The form of the commutators
 (\ref{su3canonical}) suggests that for the 1/4 BPS case the
 localization occurs as $S^4 \rightarrow S^3$. We will confirm
 this intuition by comparing with the dual string states. But first
 lets try to match the canonical structure as we did with the 1/2
 BPS condensates.

It is now important to remind the reader that the canonical
structure found in the string theory side is sensitive to 1) the
string configurations that we are considering and 2) the gauge
choice in the sigma model action. One could think that since the
states (\ref{special14BPS}) are a restricted subset of the $SU(3)$
sector (ignoring backreaction) one will not be able to match the
canonical structure found in the gauge theory side. However this is
not correct since we can always expand around these states and, if
we take into account the backreaction to the 1/4 BPS condensates, we
have a complete basis of states for this sector\footnote{Note that
even if we include the backreaction of the condensate, all terms in
the Hamiltonian will be written in terms of the operators
(\ref{su3canonical}) and thus the canonical structure will be
unchanged.}. Thus in this case the particular canonical structure is
tied to the choice of basis and therefore, in the string theory
side, it will be related to the gauge choice in the sigma model. We
would have obtained a different canonical structure had we expanded
around 1/8 BPS condensates which are of the form $\tr(\{X^n Y^m
Z^p\}\cdots)$ for example (see below).

We can now find the sigma model action for the $SU(3)$ sector in the
``fast string" limit: $J_X \rightarrow \infty$, $\lambda /J_X^2 =
\text{fixed}$, just like we did in the $SU(2)$ sector. The form of
the operators (\ref{special14BPS}) tells us that the correct gauge
choice is the one that distributes the angular momentum in $X$
uniformly along the string. Looking at the commutators
(\ref{su3canonical}) one realizes that a convenient spacetime
coordinate system for these strings is, \bea  Z = \frac{r_1}{\sqrt{1
+ \eta}} e^{i (t + \phi_1)}\;,\;\;\;  Y = \frac{r_2}{\sqrt{1 +
\eta}} e^{i (t + \phi_2)}\;,\;\;\; X = \sqrt{\frac{\eta}{1 + \eta}}
e^{i \varphi}\;,\eea where $r_1^2 + r_2^2 = 1$. We now choose the
gauge \be t = \tau\;,\;\;\; p_\varphi = \text{const.}\ee Following
the standard procedure of eliminating time derivatives for spatial
derivatives \cite{kru2, kru3} in the sigma model one finds (see
Appendix B), \bea \label{su3action} S &=& L \int d\tau \int_0^1
d\sigma \left[\frac{r_1^2}{\eta} \dot{\phi_1} + \frac{r_2^2}{\eta}
\dot{\phi_2} \right.\nonumber \\
&& \left.- 1- \frac{\lambda\, \eta }{2(4 \pi)^2 L^2(1 +
\eta)^2}\left(\frac{\eta'^2}{4 \eta} + \sum_{\alpha =
1,2}(r_\alpha'^2 + r_\alpha^2 \phi_\alpha'^2)\right) + {\cal
O}\left(\frac{\lambda^2}{L^4}\right) \right]\;.\eea

We see that the canonical structure is exactly as in the matrix
model calculation: $p_{\phi_\alpha} \sim r_\alpha^2/\eta$ with the
other momenta set to zero. In fact, one can confirm the continuum
version of the commutators (\ref{su3commutator}) -
(\ref{su3commutator3}) by using the Dirac brackets with the
following constraints: \be f_1 = r_1^2 + r_2^2 - 1\;,\;\; f_2 =
p_{r_1}\;,\;\; f_3 = p_{r_2} \;,\;\; f_5 = p_{\phi_1} - L
r_1^2/\eta\;,\;\; f_6 = p_{\phi_2} - L r_2^2/\eta\;. \ee

The localization on the $S^3$ can be understood in the same way as
with the 1/2 BPS condensates: it correspond to the classical limit
 $\eta \rightarrow 0$.

 Note however,  that unlike in the $SU(2)$ sector, the classical Hamiltonian that follows from the
 action (\ref{su3action}) does not match with the naive classical
 limit of the matrix model Hamiltonian (\ref{su3Ham}) at one loop.
 This is indeed not surprising since we are ignoring the
 backreaction to the condensates which is unavoidable in the $SU(3)$ sector. We will come back to this point
 in section 5.

\subsection{$SU(3)$ Giant Magnons?}
A natural question to ask at this point is whether we can match the
matrix model Hamiltonian (\ref{su3Ham}) at finite $L$ with some sort
of $SU(3)$ giant magnon solution in the string theory side. In
general one would expect that in the limit $n_l + m_l \rightarrow
\infty$ each $X$ string bit would correspond to a giant magnon
configuration connecting two null geodesics of the form, \be Z = r_1
e^{i  t}\;,\;\; Y = r_2 e^{i t}\;,\;\; r_1^2  + r_2^2  = 1\;,\ee one
for each condensate to the left and right of the string
bit\footnote{We can consider the more general null trajectories $Z =
r_1 e^{i \omega_1 t}, Y  = r_2 e^{i \omega_2 t}$ with
$r_1^2\omega_1^2 + r_2^2 \omega_2^2 = 1$, but this is the same as a
redefinition of $r_i$. Therefore we will set $\omega_i = 1$ without
loss of generality.}.

Note however that these 1/4 BPS condensates do not correspond to the
Giant Magnons with multiple angular momenta studied recently in the
literature \cite{new1,new2,new3} which represent {\it bound states}
in the operator language.

The states (\ref{special14BPS}) are very restricted if we ignore the
backreaction to the condensates. Therefore, we can only expect a
matching for some very special configurations. In particular one
intuitively expects that configurations for which the ends of the
giant magnon are at different radii $r_\alpha$ are unstable and thus
would require the understanding of the backreaction since on the
dual gauge theory, one would have $Z$ and $Y$ fields flowing from
one condensate to the next. We should then consider configurations
for which $r_\alpha$ is the same at each site. In the dual string
theory this would correspond to giant magnons connecting the same
null trajectory.

The classical limit of the energy formula (\ref{su3Ham}) for these
special states is, \be \label{Esb} E_\text{string bit} =
\frac{\sqrt{\lambda}}{\pi} \sqrt{r_1^2
\sin^2\left(\frac{\Delta\phi_1}{2}\right)+r_2^2
\sin^2\left(\frac{\Delta\phi_2}{2}\right)}\;.\ee It turns out that
with our simple ansatz for the classical string (see below), the
matching with this quadratic formula only works for a subset of
these states: those with  $\Delta \phi_1 = \Delta \phi_2$. These can
be considered as ``rotations" of the usual $SU(2)$ giant magnons.
This restricted matching is hardly surprising since only in this
case the spacetime probed by the string is actually flat.  More
generally, the backreaction to the string bit should take into
account the curvature of the sphere and correct the naive square
root form (\ref{Esb}).

Let us now turn our attention to the classical string theory. We
will consider strings moving on $\mathbb{R}\times S^4$ but for
simplicity we restrict the motion on the $S^3 \subset S^4$ as
follows: \be Z = r_1 \sin[\psi(\sigma)] e^{i  (\tau +
\phi_1(\sigma))}\;,\;\; Y = r_2 \sin[\psi(\sigma)] e^{i (\tau +
\phi_2(\sigma)) }\;,\;\; X = \cos[\psi(\sigma)]\;.\ee

 After defining new coordinates $\phi_1  = \phi_+ + \phi_-$ and $\phi_2  = \phi_+ - \phi_-$,
 the Nambu-Goto action in the static gauge $t = \tau$ becomes,
 \be S_\text{NG} =
\frac{\sqrt{\lambda}}{2\pi} \int d\tau d\phi_+ \sqrt{x'^2 + x^2 \
\left[1 + 2 a \phi_-' + \phi_-'^2(1 - b x^2)\right]} \;,\ee where $x
= \sin \psi$, $a = r_1^2 - r_2^2$, $b = 4 r_1^2 r_2^2$ and the
derivative is with respect to $\phi_+$. The equations of motion are,
\bea \phi_-' &=& \frac{\alpha G - a x^2}{x^2 (1 - b x^2)}\;,
\\
\label{eom}
 u(x) G \frac{d}{dx} \left( \frac{u(x)}{G}\right) &=&  x\left[ 1 + a \phi_-' + \phi_-'^2 (1 - 2 b x^2)\right]\;,\eea
where \be G = x \sqrt{\frac{a^2 x^2 - (u^2 + x^2)(1 - b x^2)
}{\alpha^2 - x^2(1 - b x^2)}}\;,\ee  $u = dx/d\phi_+$ and $\alpha$
is an integration constant. Therefore we can reduce the problem to a
non-linear ODE for $u(x)$ .

Let us now consider the special case of $r_1 = r_2 = 1/\sqrt{2}$.
One can easily show that the reality of $G$ implies that $\alpha =
0$ and so $\phi_-' = 0$. The equations of motion reduce exactly to
the ones for the $SU(2)$ giant magnon. The energy is
\cite{diegomalda}, \be E |_{r_1 = r_2} = \frac{\sqrt \lambda}{\pi}
\left|\sin\left(\frac{\Delta\phi_+}{2}\right)\right|\;,\ee where
$\Delta \phi_1 = \Delta \phi_2 = \Delta \phi_+$.

One can also show that setting $\alpha = 0$ implies $r_1 = r_2 =
1/\sqrt{2}$. This follows from the reality of $u(x)$. Setting
$\alpha = 0$ and integrating the resulting equations of motion gives
\be u(x) = |b| \frac{x^2(1 - x^2)}{1- b x^2 }  \sqrt{\frac{1 - b
x_\text{min}^2}{x_\text{min}^2(1-x_\text{min}^2)} -  \frac{1 - b
x^2}{x^2(1-x^2)}}\;,\ee where $x_\text{min}$ is the turning point.
At the boundary $x = 1$, the reality of $u(x)$ requires $b = 1$
which in turn implies $r_1 = r_2 = 1/\sqrt{2}$.

Now lets study the solutions with $\alpha \neq 0$. For these
strings, the equation of motion for $u(x)$ is highly complicated.
However it turns out that it has a very simple solution: the Giant
Magnon of \cite{diegomalda}. To see this, we first note that the
boundary condition $u(x = 1)$ is actually determined by the EOM.
When we set $x = 1$ in (\ref{eom}) all dependence on $u'(x)$ drops
and one is left with an equation for $u(1)$: \be \label{u1} u(1) =
\frac{1}{\alpha} \sqrt{a^2 - \alpha^2}\;,\ee where we need
$\text{sign}(\alpha) = \text{sign}(a)$, and $a^2 \geq \alpha^2$.
This last inequality also follows from the reality of $G$ at the
boundary $x = 1$.

We can now relate this to the Giant Magnon solution in
\cite{diegomalda} for which, \be \label{uxmalda} u(x) =
\frac{x^2}{x_\text{min}} \sqrt{1 - \left(
\frac{x_\text{min}}{x}\right)^2}\;.\ee Comparing (\ref{uxmalda})
with (\ref{u1}) we find that $\alpha = a x_\text{min}$. With
$\alpha$ now determined by the turning point $x_\text{min}$ we can
easily check that (\ref{uxmalda}) satisfies (\ref{eom}).  Therefore,
as pointed out before, the solutions that we have found are just
rotations of the Giant Magnon of \cite{diegomalda}. Nevertheless
they are consistent with the interpretation in terms of 1/4 BPS
condensates. This is because the ends of the Giant Magnon travel
along null geodesics in $S^3$ which carry two angular momenta
corresponding to the two fields $Z$ and $Y$ in the condensate, which
is itself a rotation of the 1/2 BPS condensate.

The generalization to 1/8 BPS condensates should be obvious by now.
These will be of the form $\tr(\{X^n Y^m Z^p\}\cdots)$ but since
there are no transverse excitations left, the inclusion of the
backreaction is unavoidable.  The eigenvalue distribution turns into
a singular $S^5 \subset \mathbb{R}^6$ with radius $r_0 =
\sqrt{N}/2$. In this case it is perhaps more useful to use a $SO(6)$
invariant notation as in \cite{dds2}. In this case, the dispersion
relation in any direction $a = 1,\ldots, 6$ is, \be w^a_{i j} =
\sqrt{1 +\frac{\lambda}{(2\pi)^2} |\vec x_i - \vec x_j|^2}\;,\ee
where $\vec x_i^2 = 1$ are the coordinates in the $S^5$.

\section{Backreaction to BPS Condensates}
To simplify the discussion we can look at the 1/4 BPS condensates of
the $SU(2)$ sector. That is consider operators of the form \be
\label{su214bps}\tr(Y \{Z^{n_1} Y^{m_1}\} Z \{Z^{n_2} Y^{m_2}\}
\cdots)\;.\ee  This way we can work with the simpler matrix model
(\ref{action}). The proposal is that we consider expanding around
the ``classical" configuration of commuting matrices $[Y, Z] = [Z,
\bar Z] = [Y , \bar Y]= 0$ in the action (\ref{action}) as we did
for the $SU(3)$ sector. Then the two possible excitations $Y$ and
$Z$ outside the condensates will be described by the backreaction
terms $\delta Y$ and $\delta Z$ in the matrix model.

In this case it is difficult to make comparisons with the string
theory dual because, as we will see, the number of excitations
outside the condensates is not conserved. We can, however try to
match the qualitative picture we expect from a formal field theory
calculation using the operators (\ref{su214bps}). In the limit,
$n_i, m_i \rightarrow \infty$ we again expect that the ``impurities"
outside the condensates will not interact with each other. On the
other hand, we expect only interactions between impurities and the
condensates. First, lets consider the quadratic fluctuation around
the commuting background.

Expanding $Z_\alpha  \rightarrow Z_\alpha + Y_\alpha$ where
$Z_\alpha$ is  the commuting background, one finds the following
quadratic Hamiltonian (after diagonalizing the background) \bea
H^{(2)} = \sum_{i,j} (\pi_\alpha)_i^j (\bar \pi_\alpha)_j^i +
M^{\alpha \beta}_{i j} (Y_\alpha)_i^j (\bar Y_\beta)_j^i \;, \eea
where, \be M^{\alpha \beta}_{i j} =  \left(
  \begin{array}{cc}
    1 + \alpha  |  z_{i j}|^2 \;\;\;\; & -\alpha   z_{i j} \bar y_{i j} \\
    -\alpha   \bar z_{i j}  y_{i j} \;\;\;\; &  1 + \alpha |  y_{i j}|^2\\
  \end{array}
\right) \ee and $\alpha = \lambda/(2 \pi)^2$, $z_{i j} = z_i - z_j$
and similarly for $y$.

Naively one might think that the mass matrix can be diagonalized.
However, we need to be very careful with the gauge invariance of the
states. Diagonalizing the mass matrix means that we make a change of
basis that depend on the background. On the other hand, any change
of basis must be of the form, \be (Y_\alpha)_i^j = \sum_\beta
f^\beta_{i j} (\phi_\beta)_i^j\;,\ee where $\phi_\beta$ are the
(normalized vectors) that diagonalize the mass matrix and, by gauge
invariance,  $f^\beta_{i j}$ must only depend on positive powers of
$ z_{i j}$ and $ y_{i j}$ (and their conjugates). Diagonalizing the
mass matrix one finds the following eigenvectors and corresponding
masses: \bea (\phi_1)_i^j &=& \frac{1}{\sqrt{ | z_{i j}|^2 + | y_{i
j}|^2}} \left(
                                           \begin{array}{c}
                                              z_{i j} \\
                                             - y_{i j} \\
                                           \end{array}
                                         \right)\;, \;\;\; M_1^2 = 1 + \alpha (| z_{i j}|^2 + | y_{i j}|^2) \\
(\phi_2)_i^j &=& \frac{1}{\sqrt{ | z_{i j}|^2 + | y_{i j}|^2}}
\left(
                                           \begin{array}{c}
                                              \bar y_{i j} \\
                                              \bar z_{i j} \\
                                           \end{array}
                                         \right)\;, \;\;\; M_2^2 = 1
                                         \eea

Inverting these relations one finds for example, \be (Y_1)_i^j =
\frac{1}{\sqrt{ | z_{i j}|^2 + | y_{i j}|^2}} (\bar z_{i j}
(\phi_1)_i^j+ y_{i j} (\phi_2)_i^j)\;, \ee which is not allowed by
gauge invariance. Even if we try to avoid this by not normalizing
the eigenvectors one always runs into an ill defined square root at
some point of the procedure (when defining the oscillator
operators). This is telling us that $H^{(2)}$ is really an
interacting Hamiltonian.

Using the usual oscillator basis we find $H^{(2)} = H^{(2)}_0 +
H^{(2)}_{\text int}$, where \bea H_0^{(2)} &=& w_{i j}^\alpha
(A_\alpha^\dagger)_i^j (A_\alpha)_j^i  \;,
\\H_{\text int}^{(2)} &=&  - \frac{\alpha z_{i j} \bar y_{i j} }{2 \sqrt{ w_{i j}^y w_{i
j}^z}}  (A^\dagger_y)_i^j (A_z)_j^i  + h.c. \;,\eea where  $w_{i
j}^y = \sqrt{1 + \alpha |z_{i j}|^2}$ etc. and we are taking
expectation values on holomorphic states. Therefore, we observe that
the interaction term represents the process of interchanging an
``impurity" outside the condensate with one of the fields of the
condensate (with opposite polarization): \be \tr(\cdots Z\{Z^n Y^m\}
\cdots) \leftrightarrow \tr(\cdots Y\{Z^{n+1} Y^{m-1}\}
\cdots)\;.\ee

There are also cubic and quartic interactions. Lets consider the
cubic ones. These do not preserve the number of impurities. On
holomorphic states these interactions take the following form: \be
H^{(3)} = \frac{\alpha z_{ij}}{r_0} (\tilde A_Y^\dagger)_i^j [
\tilde A_Y, \tilde A_Z]_j^i + \frac{\alpha y_{ij}}{r_0} (\tilde
A_Z^\dagger)_i^j [ \tilde A_Z, \tilde A_Z]_j^i + \text{h.c.}\;, \ee
where for simplicity of notation we defined the rescaled operators
$(\tilde A_\alpha)_i^j  = ( A_\alpha)_i^j/\sqrt{2 \omega_{i j
}^\alpha}$. Furthermore, we have assumed normal ordering of the
operators and the extra $1/r_0$ will be canceled since we loose/gain
an extra field in the operator.

 We see that these interactions involve the
absorption/emission of one impurity from the condensate. For
example, \be \tr(\cdots Z\{Z^n Y^m\}Y \cdots) \leftrightarrow
\tr(\cdots Z \{Z^{n} Y^{m+1} \} \cdots)\;.\ee Note, however, that
the interaction involves {\it both} fields $Z$ and $Y$ at each side
of the condensate. This prevents the creation of a single impurity
out of the vacuum: \be \tr(\{Z^n Y^m\}) \rightarrow \tr( \{Z^{n}
Y^{m-1}\} Y)\;.\ee This matches our intuition from the Bethe Ansatz
since there we must have zero total momentum along the trace (by
cyclicity) and therefore we need at least two Bethe roots. Moreover,
we expect that turning on a single commutator gives zero by the
cyclicity of the trace: \be \tr(\{Z^n Y^m\}) \rightarrow \tr([
\{Z^{n} Y^{m-1}\}, Y]) = 0\;.\ee

Finally we have the quartic vertex that involve the (long range)
interaction between the two impurities at each side of a condensate:
\be H^{(4)} = 2 \alpha  \tr [\tilde A_Z^\dagger, \tilde
A_Y^\dagger][\tilde A_Y, \tilde A_Z] \;,\ee where we have included
the additional $N$ that comes from the extra close loop in the
Feynman diagram. Of course we expect higher commutators from
integrating out the higher modes on the sphere.

It is easy to generalize this discussion to the $SU(3)$ sector. For
these operators we cannot avoid the inclusion of the backreaction
and of the additional interactions from the D-terms. Therefore to
fully understand the $SU(3)$ sector we need to develop new
techniques that can deal with lattices with varying number of sites.
Note that this problem is already familiar in the study of Giant
Gravitons \cite{dds3} and multiple trace operators in SYM
\cite{beisertdynamic}. We do not know how to do this at this moment.
But once this is understood we could calculate quantum corrections
to the special states studied in the previous section.

\subsection{Higher Interactions and the Strong Coupling Limit}

Now that we have some experience in defining the expansion around
commuting BPS condensates we would like to explain why this is in
fact a strong coupling expansion and when does it breaks down. In
other words, we want to understand under what circumstances, the
quandratic (or cubic) approximation is a good one.

The secret to answer this question lies in the form of the
creation/annihilation basis defined above. Note that the relation
between the creation/annihilation operators and the matrix model
coordinates is, \be (Y_\alpha)_i^j  = \frac{1}{\sqrt{2 w_{i
j}^\alpha}} [(A_\alpha^\dagger)_i^j + (\bar A_\alpha)_i^j] \sim
{\cal O}\left(\frac{1}{(1 + \lambda |\vec{x}_i -
\vec{x}_j|^2)^{1/4}} \right)\;.\ee

 Therefore, if we consider excitations that
join two eigenvalues whose distance $|\vec{x}_i - \vec{x}_j|^2$ is
fixed in the limit $\lambda \rightarrow \infty$, then all
interactions involving higher powers of the matrix model coordinates
will be naturally suppressed as, \be \tr ( Y^n) \sim
\frac{1}{\lambda^{n/4}}\;.\ee Of course, we only expect higher
commutators so we do not correct the quadratic potential. Moreover,
at least for the $SU(2)$ sector, the higher interactions must be
constrained so we do not spoil the quadratic dispersion relation.
For example, suppose we had a higher interaction term like $
 c_1 g_\text{YM}^4 \tr|[Z,[Z, Y]]|^2$. This would modify our dispersion
relation as, \be \omega_{i j} = \sqrt{1 + \frac{\lambda}{(2\pi)^2 }
|z_i - z_j|^2 + c_1 \lambda^2 |z_i - z_j|^4} \rightarrow \sqrt{1 +
\frac{\lambda}{\pi^2}\sin^2\left(\frac{p}{2}\right) + 16 c_1
\lambda^2\sin^4\left(\frac{p}{2}\right)}\;.\ee

Now, note that keeping fixed the distance between the eigenvalues is
just the Hofman-Maldacena limit \cite{diegomalda} and it was the
limit studied above. As an example, consider the quartic interaction
$V_4 \sim g^2_{YM} \tr|[Y_\alpha , Y_\beta]|^2$. It is easy to see
that this will naturally be of ${\cal O}(1)$ under this limit, while
the quadratic and cubic interactions are of ${\cal
O}(\sqrt{\lambda})$. It would be interesting to compute the one-loop
correction to the matrix model (\ref{action}) for the $SU(2)$ sector
and verify that the new interaction is indeed suppressed.

On the other hand, we can try to define the more familiar BMN limit
using the matrix model. In this case we need to take $\lambda
|\vec{x}_i - \vec{x}_j|^2 = \text{fixed} \ll 1$. This is the limit
where we consider short string bits first and {\it then} take the
$\lambda \rightarrow \infty$ limit. For example, in the $SU(2)$
sector we first take $L \rightarrow \infty$ first and so,
$|\vec{x}_i - \vec{x}_j|^2 \rightarrow |\partial_\sigma
\vec{x}|^2/L^2 \sim {\cal O}(1/L^2)$. We see that in this limit
every higher commutator interaction to the matrix model will be
relevant and thus our quadratic approximation is invalidated. This
explains why our model is so much different from the usual one-loop
spin chain. It is because our model is well defined in the {\it
opposite} limit.

Nevertheless, the lack of interactions between the $Y$ impurities in
the $SU(2)$ model makes it possible to match the string theory
result at one loop in $\lambda$ as we did in section 3. At higher
loops we do not know if the matching requires corrections to the
quadratic Hamiltonian.

What about $1/n$ corrections? One can hope to get a better
understanding of these in the $SU(2)$ sector. For that we need to
take into account the backreaction $\delta Z$ in the matrix model.
However it can be that these corrections are entangled with the
$1/\sqrt{\lambda}$ corrections. In any case, expanding around the
normal matrix background should be regarded as an asymptotic
expansion valid for BPS condensates with large angular momentum. It
would be interesting to study this issue further.

\section{Discussion}

In this article we have attempted to clarify the relation between
the reduced matrix model approach to calculate anomalous dimensions,
and the usual operator mixing problem. This was done in terms of
what we called ``BPS condensates". These are long words in the
scalar operators that look like sections of an otherwise BPS
operator. In the matrix model these condensates were interpreted as
a classical background which we use to define the perturbative
expansion. In the dual string theory they represent (in a particular
gauge) a infinitesimal section of  a string with large angular
momenta that localizes on a null geodesic on the  $S^5$. Our method
of expanding around a background of normal commuting matrices turns
out to be a good approximation in the limit of large angular
momentum on the sphere.

This interpretation allow us to match some well known string theory
results in the limit where the condensates carry infinite angular
momentum. For the $SU(2)$ sector we were able to match the sigma
model Hamiltonian of the dual string and its canonical structure
with the matrix model Hamiltonian of the quadratic fluctuations
around the 1/2 BPS condensates. This was done in the limit where we
have giant magnons \cite{diegomalda} and fast rotating strings
\cite{stringreview}. The constrained canonical structure found in
the matrix model also made clear why the infinite momentum limit
correspond to a localization of the string on the equatorial $S^1$
of the $S^5$. Corrections to this limit are harder to understand
since they require an understanding of the backreaction of the BPS
condensates and perhaps the $1/\sqrt{\lambda}$ corrections.

For the $SU(3)$ sector the matching was limited by the fact that we
need to understand the backreaction to the BPS condensates in the
matrix model even in the limit of infinite angular momentum.
Nevertheless, we were able to match the canonical structure and the
presence of $SU(3)$ giant magnons that are in a a sense ``rotations"
of the usual $SU(2)$ giant magnon. Finally, we explained why the
reduced matrix model is more naturally defined in the strong
coupling limit of the gauge theory.

So far, significant evidence has been accumulated that the effective
Hamiltonian for states of ${\cal N} = 4$ on $\mathbb{R}\times S^3$
dual to holomorphic scalar operators on $\mathbb{R}^4$ is described
by a reduced model of matrix quantum mechanics \cite{toymodel, LLM,
berens, dds2,diego, davidcota,diegomalda}.  What is really needed at
this moment is a formal {\it derivation} of the matrix model at
strong coupling. We believe that the secret lies in expanding around
(nearly) highly supersymmetric states. This is where the BPS
condensates can be very useful, specially in the infinite momentum
limit. For example, for 1/2 BPS condensates of infinite angular
momentum, we can focus on a single transverse excitation $Y$ in an
infinite ``sea" of $Z$ fields: ${\cal O} \sim \cdots
ZZZZYZZZZ\cdots$. The Feynman diagrammatics should greatly simplify
by the fact that if the $Y$ was changed for a $Z$ field, all the
diagrams must add to zero by supersymmetry. Of course, it would be
nice to derive the matrix model without resorting to the usual
diagrammatic calculations but instead integrating out higher
spherical harmonics on the $S^3$.

Understanding the operator mixing problem in terms of a reduced
matrix model can be used as an alternative route to using
integrability in testing the AdS/CFT correspondence. This is
because, as we saw, one can match directly the Hamiltonian of the
dual string {\it and} its canonical structure instead of having to
find its spectrum. Nevertheless it would be interesting to
understand the emergence of integrable structures in the language of
the reduced matrix model. In fact, it is very useful to calculate
the scattering phase for the string bits in the $SU(2)$ sector with
the quadratic Hamiltonian (\ref{H2su2}).  Perhaps one could match
the phase calculated in \cite{diegomalda} using the sine-Gordon
model in certain limit. This will probably require the understanding
of the $1/n$ corrections in the matrix model.

\section*{Acknowledgements}
I am very grateful to my advisor prof. David Berenstein for all his
insights and his help during this project. This work was supported
in part by a DOE grant DE-FG01-91ER40618.

\appendix
\section{Anti-Normal Ordering}
In this section we prove the identity, \be \int
\frac{d\Omega_3}{\text{Vol}(S^3)} \psi_{n_1, n_2}(Y,Z)^* Z^n \bar
Z^m Y^k \bar Y^l \psi_{n_1', n_2'}(Y,Z)  = \bra n_1, n_2| \anti
(\hat Z^\dagger)^n \hat Z^m (\hat Y^\dagger)^k \hat Y^l \anti | n_1'
, n_2'\ket\;,\ee where the operators in the RHS were defined in
(\ref{su3canonical}). Now, we can always identify the result of the
integration with an effective operator, \be \label{op} [Z^{n_1} \bar
Z^{m_1} Y^{n_2} \bar Y^{m_2}]_\text{int}| \tilde n_1, \tilde n_2\ket
\simeq \frac{c_{\tilde n_1 + n_1 - m_1, \tilde n_2 + n_2 - m_2}
c_{\tilde n_1, \tilde n_2}}{(c_{\tilde n_1+ n_1, \tilde n_2 +
n_2})^2} | \tilde n_1 + n_1 - m_1, \tilde n_2 + n_2 - m_2\ket \;,\ee
where, \be c_{n,m} = \sqrt\frac{(2 + n + m)!}{2 n! m!}\;.\ee

All we need to do now is to match the RHS of (\ref{op}) with the
result of the anti-normal ordered form of the dual operators
(\ref{su3canonical}). For simplicity we will do this only in the
case where $n_i - m_i >0$. The other cases follow similarly. For the
integrations we will use the identity, \be \label{id} c_{\tilde n_1
+ n_1, \tilde n_2 + n_2} = c_{\tilde n_1, \tilde n_2}
\prod_{k=1}^{n_1} \frac{1}{\sqrt{ \tilde n_1 + k}}\prod_{l=1}^{n_2}
\frac{1}{\sqrt{ \tilde n_2+ l}}\prod_{s = 1}^{n_1 + n_2} \sqrt{2 +
\tilde n_1 + \tilde n_2 + s} \;.\ee Therefore, we can write the RHS
of (\ref{op}) as, \bea \frac{c_{\tilde n_1 + n_1 - m_1, \tilde n_2 +
n_2 - m_2} c_{\tilde n_1, \tilde n_2}}{(c_{\tilde n_1+ n_1, \tilde
n_2 + n_2})^2} &=& \prod_{k = 1}^{n_1 - m_1} \sqrt{\tilde n_1 + k}
\prod_{k = n_1 - m_1  +1}^{n_1} (\tilde n_1 + k) \nonumber
\\
&& \times \prod_{l = 1}^{n_2 - m_2} \sqrt{\tilde n_2 + l} \prod_{l =
n_2 - m_2 + 1}^{n_2} (\tilde n_2 + l) \nonumber \\
&& \times \prod_{s = 1}^{n_1 - m_1 + n_2 - m_2} \frac{1}{\sqrt{2 +
\tilde n_1 + \tilde n_2 + s}} \nonumber\\
&&\times \prod_{s = n_1 - m_1 + n_2 - m_2 + 1}^{n_1 + n_2}
\frac{1}{2 + \tilde n_1 + \tilde n_2 + s}\;.\eea

It is now straightforward to verify that one gets the same result
using the anti-normal ordering form of the operators
(\ref{su3canonical}): $\hat Z^{m_1} \hat Y^{m_2} (\hat
Z^{\dagger})^{n_1} (\hat Y^\dagger)^{n_2} |\tilde n_1, \tilde
n_2\ket$ in the case $n_i - m_i > 0$.

\section{$SU(3)$ String Action}
Here we derive the action for fast rotating strings in the $SU(3)$
sector (\ref{su3action}). As usual, we start with the Polyakov
action in momentum space \cite{kru2}, \be \label{Sp} S_p =
\frac{\sqrt \lambda}{4\pi} \int d\tau \int_0^{2\pi} d\sigma
\left(p_\mu
\partial_0 x^\mu + \frac{1}{2} A^{-1} \left[ G^{\mu \nu} p_\mu p_\nu
+ G_{\mu \nu} \partial_1 x^\mu
\partial_1 x^\nu \right] + B A^{-1} p_\mu \partial_1 x^\mu\right)\;,
\ee where $A=\sqrt{-g} g^{00}$, $B=\sqrt{-g} g^{01}$ and $g^{ab}$ is
the worldsheet metric.

We now consider string moving in $\mathbb{R} \times S^5$ with the
following parametrization: \be t = \tau\;,\;\;\; X =
\sqrt{\frac{\eta}{1+ \eta}} e^{i \varphi } \;, \;\;\; Y =
\frac{r_2}{\sqrt{1 + \eta}} e^{i(\tau + \phi_2)}\;,\;\;\;Z =
\frac{r_1}{\sqrt{1 + \eta}} e^{i(\tau + \phi_1)}\;.\ee The metric in
these coordinates read, \be ds^2 =  - \frac{\eta}{1  + \eta} dt^2 +
\frac{1}{1 + \eta} \left[2 dt(r_1^2 d\phi_1 + r_2^2 d\phi_2) +
\frac{d\eta^2}{4\eta} + \sum_{\alpha = 1,2} (dr_\alpha^2 +
r_\alpha^2 d\phi_\alpha^2) + \eta d\varphi^2\right]\;.\ee

We want to use the remaining gauge freedom to distribute the angular
momentum in $\varphi$ uniformly along the string. This is
appropriate to compare with the operators (\ref{su3basis1}). We have
that, \be L =\frac{\sqrt\lambda}{4\pi} \int_0^{2\pi}d\sigma
p_\varphi = \frac{\sqrt\lambda}{2} p_\varphi\;.\ee We want to expand
the action at first non-trivial order at large $p_\varphi$.

The Virasoro constraints that follow from (\ref{Sp}) are, \bea
\label{vir1}
G^{\mu \nu}p_\mu p_\nu + G_{\mu \nu} x'^\mu x'^\nu &=& 0\;, \\
\label{vir2} p_\mu x'^\mu &=&0\;.\eea We can now solve for $p_t$
using (\ref{vir1}), \be p_t = p_{\phi_1} + p_{\phi_2} -
\sqrt{\Lambda + G_{i j }x'^i x'^j + \left(\frac{1 +
\eta}{\eta}\right) p_\varphi^2}\;,\ee where, \be \Lambda
=(p_{\phi_1} + p_{\phi_2})^2 + G^{i j }p_i p_j + \left(\frac{\eta}{1
+ \eta}\right) \frac{p_i x'^i}{p_\varphi^2}\;,\ee and $i, j  = \eta,
r_1, r_2 ,\phi_1, \phi_2$ and we have used (\ref{vir2}) to solve for
$\varphi'$.

We can now plug the value of $p_t$ calculated above back into the
action. We get an effective action in terms of the momenta $p_i$.
Since the momenta enter only algebraically into the action, we can
easily solve for them using their equations of motion. Plugging the
result back into the action we get, \be \label{Spfinal} S_p = -
\frac{\sqrt\lambda}{4 \pi} \int d\tau \int_0^{2\pi} d\sigma
\sqrt{\left( G_{i j} x'^i x'^j + \frac{1 + \eta}{\eta}
p_\varphi^2\right)\left(1
 - g_{i j } \dot x^i \dot x^j \right)}\;,\ee
where, \be g_{i j } = \frac{1}{2} \frac{\partial^2 \Lambda}{\partial
p_i
\partial p_j} \;,\ee
and $\dot x^{\phi_\alpha} \equiv 1 + \dot \phi_\alpha$.

One can make a systematic expansion at large $p_\varphi$ where one
gets an effective action which is linear in the $\dot x^i$ and one
eliminates higher powers of the time derivatives in terms  of higher
spatial derivatives \cite{kru2, kru3}. Here we do not need to follow
this procedure in detail since we want the leading order at large
$p_\varphi$. Therefore, as usual we assume that all time derivatives
are of order $\sim 1/p_\varphi^2$. Then expanding the action
(\ref{Spfinal}) at leading non-trivial order we find the result
(\ref{su3action}). Higher  order corrections will only affect the
form of the Hamiltonian but not the canonical structure.

\end{document}